\documentclass[aps,superscriptaddress,amsfonts,amssymb,amsmath,showpacs,nofootinbib,twocolumn]{revtex4-1}
\usepackage[dvips]{graphicx}

\begin{document}

\title{Black hole hair in Lovelock gravity}

\author{Jozef Sk\'akala}
\email{email: jozef@iisertvm.ac.in}
\affiliation{School of Physics, Indian Institute of Science, Education and Research (IISER-TVM), Trivandrum 695016, India}

\author{S. Shankaranarayanan} 
\email{e-mail: shanki@iisertvm.ac.in} 
\affiliation{School of Physics, Indian Institute of Science, Education and Research (IISER-TVM), Trivandrum 695016, India}

\begin{abstract}
We present a method to analyse black hole hair in the spherical symmetric sector of the Lovelock theory 
in arbitrary dimensions that is an alternative to solving the equations of motion in their complete form. 
We explicitly show that the method matches with the known black hole solutions for the vacuum and electro-vacuum spacetimes in 
Lovelock gravity theories. We further apply the method to the case of minimally coupled non-self-interacting massless scalar field and prove that 
there is no (non-self-interacting) massless scalar hair for the spherically symmetric Lovelock black holes.
\end{abstract}

\maketitle

\section{Introduction}

Lovelock theories of gravity \cite{Lovelock} are an unique higher
dimensional generalization of Einstein gravity, such that the gravity
equations of motion remain second order (and quasilinear in second order). They provide 
a natural arena for understanding many deep features of gravity and 
recently they have been a subject of an intense study. (For a recent 
review see \cite{Paddy1}.) Some higher dimensional Lovelock theories arise also as  
a weak field limit of string theory \cite{Witten, Gross}.

Although the vacuum black hole solutions of the theory are reasonably
well understood \cite{Deser, Maeda, Camanho}, (at least in the spherically
symmetric sector), much less it is known about black hole solutions
that include matter fields. Understanding the solutions with matter is
very interesting from the point of view of no-hair theorem. This 
work is an attempt to identify an efficient method 
that will help to analyse black hole hair in the Lovelock gravity theories. 
We show explicitly that if one is interested only in the horizon\footnote{Obviously, 
the concept of horizon is in general observer dependent. However, for spherically 
symmetric static spacetimes, by ``horizon'' we refer to a horizon associated with 
static observers.} structure of the solution (and eventually in the horizon's thermodynamical
properties \cite{Maeda, Camanho, Myers}), {\it it is sufficient to solve
equations of a significantly reduced theory}, instead of dealing with
the dynamical equations of the Lovelock theory in their complete form.

In particular, for spherically symmetric sector one can use the 2D
dilaton gravity coupled with matter, where all the information about
Lovelock coupling constants is encoded in the dilaton potential. We
apply this idea to the case of Lovelock theory coupled with
electromagnetic (EM) field and to the case where Lovelock gravity is
minimally coupled to both, electromagnetic and non-self-interacting massless scalar
field. In the latter case we prove that there is \emph{no} massless
scalar hair for Lovelock black hole.

In the rest of this work, $D$ refers to number of spacetime dimensions 
and we set $G=c=1$.  We reserve the letter $r$ for the geometrical
quantity labelling the $(D - 2)$ spheres of spherical symmetry as
$r=\sqrt{A/4\pi}$, where $A$ is the area of the $(D - 2)$ sphere.

~

\section{Near horizon reduction of spherically symmetric Lovelock gravity to a 2D dilaton theory}

The Lovelock gravity theory (ignoring the surface term) is defined by the action of the form:
\begin{equation}\label{act}
S=\int d^{D}x\sqrt{-g}\cdot\sum_{m=0}^{[D/2]}\lambda_{m}L_{m},
\end{equation}

with

\begin{eqnarray}
L_{m}=\frac{1}{2^{m}}\delta_{c_{1}...c_{m}d_{1}...d_{m}}^{a_{1}...a_{m}b_{1}...b_{m}}
R^{c_{1}d_{1}}_{~~~~~a_{1}b_{1}}...R^{c_{m}d_{m}}_{~~~~~a_{m}b_{m}}.
\end{eqnarray}

Now consider a spherically symmetric line element:
\[ds^{2}=\gamma_{ij}dx^{i}dx^{j}+r^{2}d\Omega^{2}_{D-2},~~~~i,j=0,1.\]
It can be shown that for the spherically symmetric sector, near a spacetime
horizon, the general Lovelock action can be (up to surface terms)
reduced to the following form \cite{Cvitan}:
\begin{widetext}
\begin{eqnarray}\label{1}
I_{R}=V_{D-2}\sum_{m=0}^{[D/2]}\left\{\lambda_{m}\frac{(D-2)!(D-2m)(D-2m-1)}{(D-2m)!}\int
dx^{2}\sqrt{-\gamma}~r^{D-2m-2}\cdot\right.~~~~~~~~~~~~~~~~\\ 
\left.\left(-m(\nabla r)^{2}+\frac{m
  Rr^{2}}{(D-2m)(D-2m-1)}+1\right)\right\}.~~~~~~~\nonumber
\end{eqnarray}
\end{widetext}
Here $V_{D-2}$ is a volume of a unit $(D-2)$ sphere and $R$ is the Ricci scalar of the 2D geometry. The near-horizon
approximation uses the fact that the scalar $(\nabla r)^{2}$ vanishes
as one approaches horizon, and one can therefore neglect all the
powers of $(\nabla r)^{2}$ higher than one.

The action (\ref{1}) can be further rewritten \cite{Cvitan} through
the redefinition of variables:
\[\phi=2V_{D-2}\sum_{m=1}^{[D/2]}m\lambda_{m}\frac{(D-2)!}{(D-2m)!}r^{D-2m},\]
and a conformal transformation
\begin{equation}\label{conf}
\tilde\gamma_{ab}=\frac{d\phi}{dr}\gamma_{ab},
\end{equation}
into the following reduced action:
\begin{equation}\label{2}
I_{R}=\frac{1}{2}\int d^{2}x \sqrt{-\tilde\gamma}\left(\phi R(\tilde\gamma)+V(\phi)\right).
\end{equation}
Here the dilaton potential $V(\phi)$ is given as:
\[V(\phi)=\left(\frac{d\phi}{dr}\right)^{-1}\sum_{m=0}^{[D/2]} (D-2m-1)\tilde{\lambda}_{m}  \, r(\phi)^{D-2m-2},\]
with
\[\tilde{\lambda}_m = 2 V_{D-2} \, \frac{(D-2)!(D-2m)}{(D-2m)!} \, \lambda_m .\]
The function $r(\phi)$ is, in general, multivalued.
Hence, here (and also later) we use physical arguments to 
choose a branch that represents inverse to $\phi(r)$ on the relevant domain. 

We refer to the theory following from the action (\ref{2}) as ``effective Lovelock theory'' and abbreviate 
it as ``ELT''. The full Lovelock theory following from the action (\ref{act}) we abbreviate as 
``LT''.

\section{The information encoded by the 2D dilaton theory}

 Let us assume that one of the theories (ELT or LT) has a
solution with a horizon localized at $r_{H}$. Since the equations of ELT and LT match near the horizon, one can use this as
initial data in the equations of the other theory and one shall
locally obtain a solution of the other theory. The solution of
the other theory has by construction a horizon located at the same
$r_{H}$.

Therefore we conjecture the 
following:  Let us fix all the values of the coupling constants in LT (and 
therefore also in ELT). 
The (spherically symmetric) solutions of ELT and solutions of LT are labelled (for fixed values of coupling constants) by
one additional\footnote{There still can be different finite number of branches of
solutions.} continuous parameter $M$. Take a solution / class of solutions of the ELT with 
some fixed value of $M$, such that it is a solution / class of solutions with a
horizon. Then such solution / class of solutions corresponds to a
solution / class of solutions of LT associated with a horizon with the
same area (or $r_{H}$) than the original solution of ELT. (For the two solutions, 
the value of $M$ can, in general, be different.) The
correspondence holds also the other way round: To some
stationary\footnote{In Lovelock theories, there exist special
  degenerate non-stationary vacua, for special values of coupling constants, 
  that accommodate maximally symmetric $(D-2)$ submanifold and they provide the 
  only exception to Birkhoff theorem \cite{Maeda}.} solution/class of solutions (with $M$) of the LT with
a horizon, there corresponds a solution/class of solutions of ELT with
the same horizon, the same values of coupling constants and generally
different $M$.  (Note that we generally speak about ``class of
solutions'' as there can be multiple branches of solutions with the
same $M$. This is indeed a common feature of Lovelock theory
\cite{Maeda, Camanho}.)
 
Let us now test the above observation: The variation of the action of ELT given by Eq.(\ref{2})
gives the following dynamical equations:
\begin{eqnarray}
\label{dyn1}
R(\tilde\gamma)+\frac{dV(\phi)}{d\phi}&=&0 \\
\label{dyn2}
\nabla_{\mu}\nabla_{\nu}\phi-\frac{1}{2}g_{\mu\nu}V(\phi)&=&0 
\end{eqnarray}

It is known \cite{Cavaglia} that the most general solution
of these equations is static and can be expressed as:
\[\tilde\gamma_{ij}dx^{i}dx^{i}=-F(\phi)dt^{2}+F^{-1}(\phi)d\phi^{2},\]
with 
\begin{equation}\label{F}
F(\phi)=\int^{\phi}V(\phi')d\phi'-2M,
\end{equation}
where $M$ is some constant chosen to be the mass parameter of the
solution. (For the derivation of this result see the Appendix (\ref{A}).)

Let us explore the horizon structure of the solutions of ELT. (It is
obvious that this is the horizon structure of the solutions of the
higher dimensional reduced theory, as the conformal 2D transformation
given by Eq.(\ref{conf}) does not effect the horizon structure of the
spacetime.)  The Eq.(\ref{F}) leads to:
\begin{eqnarray}\label{solution}
F(r)=\left(\sum_{m=1}^{[D/2]}\tilde{\lambda}_{m} \, r^{D-2m-1}\right)-2M.
\end{eqnarray}

The horizons are then obtained by simple zeros of $F$:
\begin{eqnarray}\label{rootseven}
\left(\sum_{m=1}^{[D/2]}\tilde{\lambda}_{m} ~r_{H}^{D-2m-1} \right)- 2 M=0.
\end{eqnarray}

As mentioned earlier, the solutions of the \emph{vacuum}
spherically symmetric LT are reasonably well known and generally split
into different types \cite{Maeda}. However the only relevant
type are the static solutions and these are the Tangherlini-like
solutions \cite{Maeda} that can be expressed through our parameters as ($M$ is the
generalized Misner-Sharp mass):
\begin{equation}\label{Tan}
2M=\sum_{m=0}^{[D/2]}\tilde\lambda_{m}r^{D-1-2m}\left\{1-f(r)\right\}^{m}.
\end{equation}
One can obtain the horizon structure from Eq.(\ref{Tan}) by putting $f=0$ and it is easy to see that this leads precisely to our Eq.(\ref{rootseven}) derived through ELT. 
Note that this shows that ELT not only encodes the 
horizon structure of LT, but also the mass parameter of the solution of ELT 
and the generalized Misner-Sharp mass of the solution of LT \emph{relate by a simple identity}.

However, from the point of view of ELT, vacuum spacetimes are 
not interesting, as one knows all the information that ELT can
offer from an exact result of LT. (See  Eq. (\ref{Tan}).) The advantage 
of ELT becomes clear in the case of non-vacuum spacetimes. In this case, 
as we show in the rest of this work, ELT provides an efficient method 
to decode essential information about the horizon
structure of the spacetime and if the horizons are present, it also
encodes the essential information about horizon thermodynamics. (One
can calculate entropy and temperature as functions of coupling
constants, physical parameters of the fields and $M$, and further
analyse thermodynamical stability \cite{Maeda, Camanho}.) 

All the 
details about the horizon structure of a solution of some particular 
LT are encoded in the potenial $V(\phi)$ of ELT and
some diffeomorphism invariant parameters of the fields (like
electromagnetic charge for the electromagnetic (EM) field). Therefore,
as mentioned in the introduction, ELT can be a very useful tool to
investigate the \emph{hair} of Lovelock black holes. In the rest, we 
will demonstrate this by considering two specific  -- 
electromagnetic and masseless scalar --- matter fields.

\section{Using the effective theory to explore the hair of Lovelock black holes}

The first application lies in exploring the hair from the EM field and is again
more of a consistency check as the results are already known
\cite{Wiltshire}. The second result is more interesting and we will 
demonstrate that there is no hair for Lovelock black holes from the 
massless minimally coupled non-self-interacting scalar field. Although, this might be 
expected from a general argument formulated by Bekenstein \cite{Bekenstein1, 
Bekenstein2}, it is nice to offer an independent proof explicitly 
for the Lovelock theories. Note that Bekenstein's argument \cite{Bekenstein2} is very general and works for a significant class of self-interacting scalar fields, however it makes some mild technical assumptions on the boundaries, such as asymptotic flatness\footnote{However, to extend the argument of Bekenstein \cite{Bekenstein2} to asymptotically deSitter spacetime seems straightforward.}, which we do not need to make.

\subsection{The electromagnetic field}
Let us add to the action (\ref{2})
the action of the electromagnetic field with the Lagrangian:
\[L_{EM}=-\frac{1}{4}W(\phi)F^{(2)}_{\ij}F^{(2) ij}.\]
For a particular choice of $W(\phi)$:
\begin{equation}\label{W}
W(\phi)=\frac{V_{D-2}}{4\pi}r^{D-2}\frac{d\phi}{dr},
\end{equation}
one obtains the reduced theory of the minimally coupled EM field to
the $D$ dimensional Lovelock gravity.

It can be shown \cite{Kunstatter, Medved} that by redefining 
\begin{equation}\label{transformation}
V(\phi)\to V(\phi)-\frac{Q^{2}}{W(\phi)},
\end{equation}
where $Q$ is the EM charge, the solution of the coupled gravity-EM
theory can be expressed in the same form as without the EM field, as given in Eq.(\ref{F}). 
The horizon structure of the
solutions of LT-EM theory can be then obtained from ELT-EM theory in
analogy to the vacuum case as:
\begin{eqnarray}\label{rootseven2}
\sum_{m=1}^{[D/2]}\tilde{\lambda}_{m} r_{H}^{D - 2m -1} - 
2 M+\frac{4 \pi Q^{2}r_{H}^{3-D}}{(D-3)V_{D-2}}=0.~~~~
\end{eqnarray}
This is again the same result that can be derived from the full
solution of coupled LT-EM theory \cite{Wiltshire, Maeda}.

\subsection{No massless scalar hair for Lovelock black holes}

It is well known that there is no massless (non-self-interacting), minimally coupled scalar 
hair for 4-dimensional spherically symmetric general relativistic 
black hole \cite{Bekenstein1, Bekenstein2, Chase, Wyman, Virbhadra}. The most general solution describing static (non-self-interacting) massless scalar field 
coupled to 4D Einstein gravity is the so-called Wyman's solution \cite{Wyman, Virbhadra}:
\[-f^{\gamma}dt^{2}+f^{-\gamma}dx^{2}+f^{1-\gamma}x^{2}d\Omega^{2},\]
where
\[f(r)=1-\frac{2\sqrt{M^{2}+s^{2}}}{r}, \quad  \quad 
\gamma=\frac{M}{\sqrt{M^{2}+s^{2}}},\]
and $s$ is scalar field charge. It can be easily observed that $\gamma$ has values in the interval $\gamma\in (0,1]$. 

In order to have a black hole (BH) horizon, $f$ must necessarily
vanish at the horizon, but the angular part of metric has to be in the same time non-zero.  This is possible only for $\gamma=1$, such that corresponds 
to zero scalar charge $s=0$ and describes Schwarzschild BH. (Note that if the angular part of the metric is zero, one obtains a central naked singularity, not a BH.) This means, if one adds to Schwarzschild BH a static spherically symmetric configuration of a massless scalar field with an infinitesimal charge $s$, the hole disappears.

Let us now focus on the general Lovelock case. To the ELT action (\ref{1}), 
we add the matter contribution from the massless scalar field and the 
EM field:
\[L_{EM+S}=-\frac{1}{4}W(\phi)F_{ij}F^{ij}-\frac{1}{2}S(\phi)\psi_{,i}\psi^{,i}.\]
Again $S(\phi)$ is so far kept as a general function, but for the reduced
theory one has:
\begin{equation}\label{S}
S(\phi)=V_{D-2}r^{D-2}.
\end{equation}

(The function $W(\phi)$ is for the reduced theory given again by
Eq.(\ref{W}).) For convenience, as in the previous case, let us encode
the EM part of the theory in the potential $V(\phi)$ through the Eq.(\ref{transformation}).

Assuming stationarity of scalar field and consequently of
metric, the 2-D line element is given by:
\[-Fdt^{2}+\frac{\Omega^{2}}{F}dx^{2},\]
where $\Omega$ is a arbitrary (non-zero) smooth function of $x$.  
Solving the scalar field equation of motion leads to:
\begin{equation}\label{scalar}
\left(\frac{S(\phi) F \psi_{,x}}{\Omega}\right)_{,x}=0~~\to~~\psi_{,x}=\frac{\sqrt{V_{D-2}}\cdot s\cdot \Omega}{S(\phi) F}.
\end{equation}
Here $s$ is the constant of integration and is chosen in such a way 
that it matches with the scalar charge from Wyman's solution for 4-dimensional Einstein
gravity.  Eq.~(\ref{dyn1}) and Eq.~(\ref{dyn2}) take the form:
\begin{eqnarray}
\label{one}
& & -\frac{F_{,xx}}{\Omega^{2}}+\frac{\Omega_{,x}F_{,x}}{\Omega^{3}}+\frac{dV(\phi)}{d\phi}=\frac{dS}{d\phi}\cdot\frac{V_{D-2} s^{2}}{S^{2}(\phi)F}, 
\end{eqnarray}
\begin{eqnarray}
\label{two}
\frac{F F_{,x}\phi_{,x}}{2\Omega^{2}}-\frac{F^{2}\Omega_{,x}\phi_{,x}}{\Omega^{3}}+~~~~~~~~~~~~~~~~~~~~~~~~~~~~~\nonumber\\
\frac{F^{2}\phi_{,xx}}{\Omega^{2}}-\frac{F V(\phi)}{2}=\frac{V_{D-2} s^{2}}{2S(\phi)},~~ 
\end{eqnarray}

\begin{eqnarray}
\label{three}
-\frac{1}{2}\frac{F F_{,x}\phi_{,x}}{\Omega^{2}}+\frac{1}{2}FV(\phi)=\frac{V_{D-2} s^{2}}{2S(\phi)}.
\end{eqnarray}
Although the above equations have three parameters $F,\Omega,\phi$, there 
are only two real degrees of freedom. The extra parameter is a gauge ambiguity and 
can be removed by gauge fixing. It can be observed that one can always fix the 
gauge as $\Omega=1$. For such a choice of $\Omega$, Eq.~(\ref{three}) reduces to:
\[-\frac{1}{2}F F_{,x}\phi_{,x}+\frac{1}{2}FV(\phi)=\frac{V_{D-2} s^{2}}{2 S(\phi)}.\] 

In this gauge, the horizon corresponds to $F=0$. Let us make a
reasonable assumption that wherever $F$ is defined, the product
$F_{,x}\phi_{,x}$ is well defined and finite. As can be seen from 
Eq.~(\ref{S}), for the reduced theory $S(\phi)^{-1}$ is finite and non-zero on the relevant 
domain. (The relevant domain excludes an infinitesimal neighbourhood of the radial center.) Let us also assume that $V(\phi)$ is on the relevant domain non-singular. (Indeed in the following paragraph we prove that singularity of $V(\phi)$ implies curvature singularity. Also note that the naked singularity of the Wyman's solution 
($s\neq 0$) is located at the radial center and $V(\phi)$ is singular there.)  

Based on these assumptions, plugging $F=0$ in Eq.~(\ref{one}) we get:
\[\frac{V_{D-2} s^{2}}{2S(\phi)}=0~~~\to~~~ s=0.\]
Therefore the presence of the horizon implies zero scalar field charge.   

Now let us deal with the case of $V(\phi)$ being singular.  $V(\phi)$ can be rewritten as:
\[V(\phi(r))=\frac{\tilde V(\phi(r))}{d\phi/dr},\]
where $\tilde V(\phi(r))$ is a regular function of $r$.  The
singularity therefore corresponds to
\[\frac{d\phi}{dr}=0.\] 

However we require Ricci scalar to be finite at the horizon\footnote{We exclude scalar curvature singularities at the horizon. Such singularities lead to infinite tidal forces.}, even if $V(\phi)$ is singular. In such case Eq.\eqref{one} suggests that:
\begin{equation}\label{sing}
\frac{dV(\phi)}{d\phi}-\frac{dS(\phi)}{d\phi}\frac{C}{F},
\end{equation}
should be finite. (Here $C$ is some finite constant.)  One can observe
(from the assumptions) that the variables must have the
following behaviour near the horizon / singularity of $V$:
\begin{itemize}
\item $\phi_{,r}\sim z^{\alpha},~~\alpha\geq 1,$
\item $\tilde V\sim z^{\delta}, ~~0\leq\delta<\alpha,$  
\item $F\sim z^{\alpha-\delta},$
\end{itemize}
with $z=r-r_{H}$, where $r_{H}$ is the horizon radius. (Note that if $V(\phi)$ is divergent, the only relevant case to
consider is when $FV$ is finite and nonzero. Therefore
$F\sim V^{-1}$ near the horizon.)

Using the above relations, the expression in Eq.(\ref{sing}) can be shown to
behave as $\sim z^{-2\alpha-1+\delta}(\delta-\alpha)$. This leads to
the following conditions:
\[\delta\geq 2\alpha+1 \quad \mbox{or} \quad \delta=\alpha, \]
but considering the constrains on $\alpha$ and $\delta$, that is $0\leq\delta<\alpha$, none of these two conditions can be fulfilled. Therefore the singularity of $V(\phi)$ implies the singularity of $R(\tilde\gamma)$. 
This seems all very straightforward, but there is one problem: the Ricci scalar $R(\tilde\gamma)$ is not the physically relevant Ricci scalar. What remains to show is that the divergence of $R(\tilde\gamma)$ implies the divergence of the physical $D$ dimensional Ricci scalar and this will be done in the following paragraph. 

The physically relevant $D$ dimensional Ricci scalar is given as:
\begin{eqnarray*}
R^{(D)}=R(\gamma)+~~~~~~~~~~~~~~~~~~~~~~~~~~~~~~~~~~~~~~~~~~~~~~~~~~~~~~\\
+~\hbox{(terms finite everywhere outside the radial center)}.
\end{eqnarray*}
Here $R(\gamma)$ is related to $R(\tilde\gamma)$ from the Eq.(\ref{one}) through the conformal transformation:
\begin{equation}\label{conf2}
R(\gamma)=\phi_{,r}\left(R(\tilde\gamma)-2\nabla^{2}\{\ln (\phi_{,r}^{-1/2})\}\right).
\end{equation}

From the Eq.(\ref{conf2}) one can observe that near the singularity of $R(\tilde\gamma)$ there will be for $R(\gamma)$ two non-zero (potentially) divergent terms. The first term behaves as $z^{-\alpha+\delta-1}$ and is (for the relevant $\alpha,\delta$) always divergent. The other term behaves as $z^{-\delta+2\alpha-2}$ and can be in some cases divergent as well. However the divergence of the second term is for the relevant $\alpha\geq1$ always of a lower order than the divergence of the first term and therefore the divergences cannot cancel. This means the divergence of $R(\tilde\gamma)$ indeed implies divergence of the physical higher-dimensional Ricci curvature. In other words, 
the potential can be singular {\it only} if the Ricci scalar also diverges 
at the same point and there will be infinite tidal forces as one approaches the horizon.

\section{Conclusions}
In this work, we have presented an efficient and robust method to analyse 
the hair of higher dimensional Lovelock black holes. Through this method we have shown the  non-existence of massless (non-self-interacting) 
scalar minimally coupled hair for Lovelock black holes. 

It will be interesting 
to apply the method to analyse hair of more complicated matter sources (like minimally 
coupled, self-interacting scalar field). However, it is important to note that 
it may not be straightforward to do an analysis using the above method in 
case of self-interacting scalar field. Furthermore, if one derives via our method solutions with horizons, one can continue to analyse the horizon thermodynamical properties.  A particularly interesting question that can be analysed from the results obtained by our method is the thermodynamical stability of the solutions\footnote{One remarkable fact is that at least in general relativity the thermodynamical stability
carries some information about the classical dynamical stability of
the solution \cite{Wald}. However, up to the authors knowledge, this result was not yet extended beyond general relativity.}.  
All these problems are left for the future work.


\medskip

{\bf Acknowledgments:}~~ The work is supported by Max Planck partner
group in India. SS is partly supported by Ramanujan Fellowship of DST,
India.

\begin{appendix}
\section{The case for zero scalar charge \label{A}}

For the
special case of spacetime that is in higher dimension given only by
one physical degree of freedom $f(r)$ as

\[-fdt^{2}+f^{-1}dr^{2}+r^{2}d\Omega^{2}_{D-2},\]

one can choose the $\Omega^{2}$ parameter in (\ref{one}) --
(\ref{three}), as $\Omega^{2}=\phi_{,x}$. The equations (\ref{two})
and (\ref{three}) become
\begin{eqnarray}
\label{three-new1}
& & \frac{1}{2}\frac{F F_{,x}}{\phi_{,x}}-\frac{1}{2}FV(\phi)=\frac{V_{D-2}s^{2}}{2 S(\phi)}, \\
\label{three-new2}
& & -\frac{1}{2}\frac{F F_{,x}}{\phi_{,x}}+\frac{1}{2}FV(\phi)=\frac{V_{D-2}s^{2}}{2 S(\phi)}.
\end{eqnarray}
Together they lead to the condition that
\[s=0,\]
which just shows that in case of non-constant scalar field the spacetime
metric must have both physical degrees of freedom independent.  It can
be easily seen that the equations Eq. (\ref{one}) and Eq. (\ref{two})/
Eq. (\ref{three}) (which are the same) turn for the case of zero scalar charge into:

\begin{equation}
\frac{d^{2}F}{d\phi^{2}}=\frac{dV(\phi)}{d\phi},~~~\hbox{and}~~~\frac{dF}{d\phi}=V(\phi),
\end{equation}

and it is trivial to observe that the solution is given by Eq.(\ref{F}).

\end{appendix}

\end{document}